\documentstyle[11pt,paspconf]{article}

\begin{document}

\title{Bright Stars and Metallicity Spread in the Globular Cluster
$\Omega$  Centauri}

\author{Sergio Ortolani}
\affil{Department of Astronomy, Padova University}
\author{Stefano Covino}
\affil{Milano Astronomical Observatory}
\author{Giovanni Carraro}
\affil{Department of Astronomy, Padova University}

\begin{abstract}
The globular cluster $\Omega$ Centauri (NGC~5139) is the most
massive and
brightest cluster in our Galaxy. It has also a moderately high mass to
light ratio (3.6) and an anomalous flattening (0.83) for a globular  
cluster.
This cluster is also very interesting because it is one of a
few examples of globular clusters with a measurable spread in the metal
abundance (see Da Costa \& Willumsen 1981, Norris et al. 1996, and
Suntzeff and Kraft 1996 and
references therein) and then it offers a unique, big sample of nearby
stars having
all the same distance and reddening but showing different metallicity (and
age ?) effects.  A recent paper by Norris et al. (1997) shows also an
interesting correlation between kinematics and metal abundance.
\end{abstract}


\section{Introduction}
Most of the recent work has been based on
spectroscopy, while little effort was done in analyzing
the photometry.
In this poster we present new, high accuracy photometry obtained at ESO
La Silla with the New Technology Telescope ($NTT$) under good seeing
conditions ($0.7^{\prime\prime} FWHM$) in the inner
regions of the
cluster.
In particular we will focus our analysis in the field located at
$5^{\prime}$
North of the cluster center, where the crowding is considerably lower
than in the central region, but the field is still rich enough in stars to
secure more than about 1000 objects between the Main Sequence (MS) Turnoff
region
(TO) and the Horizontal Branch (HB) in a 6 square arcminute field.\\

\begin{figure}[!htb]
\plotfiddle{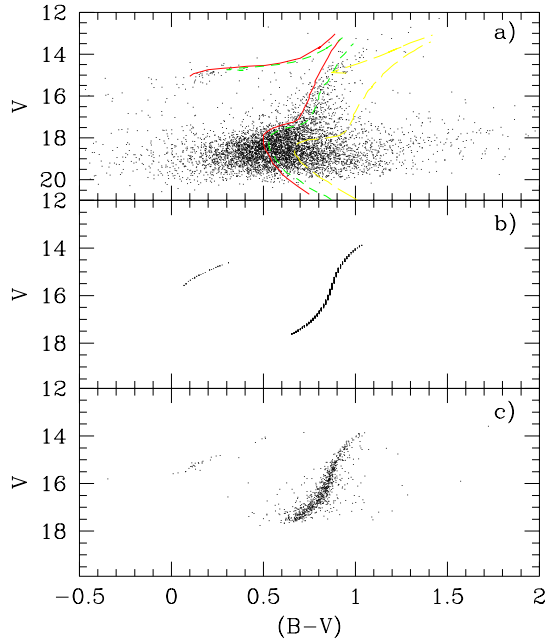}{75truemm}{0}{45}{45}{-150}{-80}
\caption{The bright portion of $\Omega$ Centauri CMD. See text for
details.}
\end{figure}

\section {The Color-Magnitude Diagram}
Fig.~1a shows the Color-Magnitude Diagram (CMD) obtained
in this field.
Overimposed are  $15~Gyr$  isochrones for  metallicities of
$z~=~0.0001$, $0.0004$ and $0.004$ (corresponding respectively to about
$[Me/H]~=~-2.3$,
$-1.7$ and $-0.7$), adopted from Bertelli et al. (1994) and shifted by
$0.12 ~mag.$
in $(B-V)$ and $14.1 ~mag.$ in $V$ to get the best fit of the $z~=~0.004$
isochrone with
the data.
The points  follow quite well the two metal poor isochrones,  
from the HB to the
TO, while the $z~=~0.04$ isochrone clearly appears at the edge of the
distribution, but still could explain the presence of a bump in the
Sub Giant Branch (SGB)
at about $V~=~14.6-14.7$, due to   metal rich HB stars superimposed.  
In order to evaluate the effects of instrumental errors on the CMD, we   
carried
out several experiments with about 1000 artificial stars injected in the
original frame, generated from the input CMD shown in Fig.~1b.
The corresponding output is shown in fig.~1c. A close analysis of the
results from the simulations indicate that the output diagram is
systematically shifted upwards, compared to the input one, as a
consequence
of blending.

The resulting spread is also higher than that obtained simply combining
the photometric errors in the two colors from single stars.

\section{Comparison with Metallicity Data}
Fig.~2a presents the histogram distribution in color of the observed data,
Fig.~2b the histogram distribution from the artificial
CMD and Fig.~2c the distribution of the SGB stars metallicities
observed
by SK without any compensation for radial and evolutionary effects.
The shape of the two histograms is surprisingly similar
in spite of the different samples used (SK sample is located at about
$10^{\prime}$
from the cluster center, while ours is at $5^{\prime}$), with a tail
toward the highest
metallicities and redder colors. 

\begin{figure}[!htb]
\plotfiddle{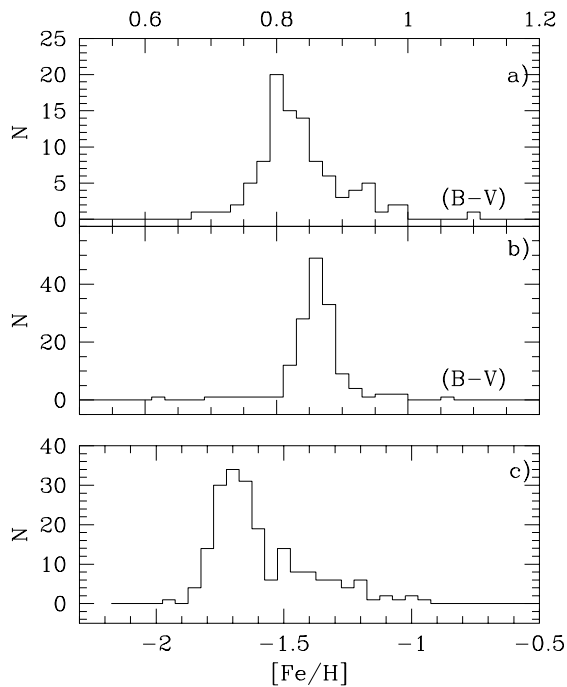}{75truemm}{0}{45}{45}{-150}{-80}
\caption{Comparison between metallicity and photometry.}
\end{figure}

The problem of the existence of a
secondary
peak due to a metal rich broad distribution discussed by Norris et al.
(1997)
cannot be confirmed by the present data but will be further investigated
after
the analysis of the full set of the available data.

Our comparison of the spectroscopic metallicities with the photometric
ones is based mainly on the intrinsic width of the SGB compared with the
width computed from the spectroscopically determined metallicity spread.
We used Suntzeff and Kraft paper (1996) as a reference because they
measured
a wide and consistent sample in the SGB. Fig.~3a has been obtained from
their
original data (Tab.~3b). It shows the relation $[Fe/H]$ versus $(B-V)$
dereddened
color (see also their fig.~9), where the high resolution metallicity scale
was
choosen. The solid line superimposed to the data is the theoretical   
relationship derived from Bertelli et al. (1994), for an age of $15 ~Gyr$,
at
the level of $0.44 ~mag$. below the HB, corresponding to SK sample of
stars.
The dashed line has been obtained using younger models ($9 ~Gyr$)
for the $[Fe/H]~=~-0.7$ isochrone. Fig.~3b is the same as fig.~3a but for
the
sample of SK giant stars located within $6.2^{\prime}$ from the cluster   
center.
Two facts are evident from these figures:\\

\begin{description}
\item{$\bullet$}~~the spread in both
metallicity
and $(B-V)$ color is very wide. At a fixed metallicity the color range
spans
about 0.2 ~mag., while at a given color the metallicity goes from
[Fe/H]=-1.8
to -1.0. This wide range is higher than the expected observational errors,
even if it is not easy to get an estimate of the photometric errors in the
colors published by SK (which are derived from Wolley, 1966);
\item{$\bullet$}~~ the
$[Fe/H]$
versus $(B-V)$ relationships are roughly in agreement with the general
trend
of the data and they get steeper for a younger age of the metal
rich component, as it is evident from the analysis of fig.~3c, where
isochrones of $15$ and $9~Gyr$ are shown, respectively.
\end{description}

\begin{figure}[!htb]
\plotfiddle{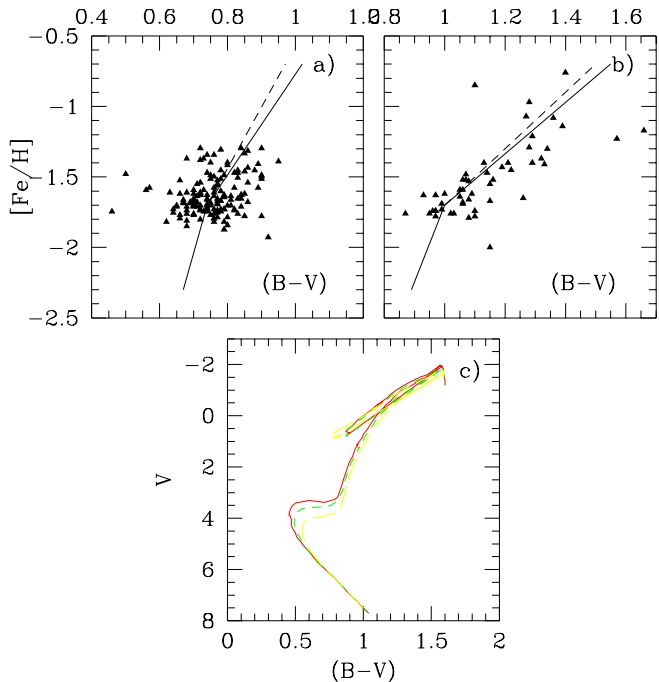}{75truemm}{0}{45}{45}{-150}{-80}
\caption{Metallicity vs dereddened colors. See text for details.}
\end{figure}

\section{Discussion}
This younger age is still compatible with the data.
An analitical, approximate expression for the intrinsic width of the SGB
can be given in the following simplified form

\[
\Delta(B-V) = \sqrt((\frac{d(B-V)}{d[Fe/H]} \times \Delta [Fe/H])^{2} +
(\frac{d(B-V)}{d\tau} \times \Delta \tau)^{2}) =
\]
\begin{equation}
           = \sqrt((0.23 \times \Delta [Fe/H])^{2} + (0.01 \times \Delta
\tau)^{2}).
\end{equation}

\noindent
where $\tau$ is the age measured in billion years.
The factor $\Delta \tau$ is negative for increasing age at a metallicity
higher than the average, positive if lower.
Using a metallicity spread $\sigma[Fe/H]~=~0.2 ~dex$ obtained by SK for 
the SGB sample,
corrected for 
radial effects, we derive a theoretical
$\sigma_{(B-V)}~=0.26 \times 0.23~=~0.06 ~mag.$
when no age difference is assumed between the metallicity components.
From our data in fig.~1 we
measured a SGB, width at $V~=~15.8$ in a half magnitude bin in
our
data obtaining $\sigma (B-V)~=~0.064 ~mag$. From the artificial CMD in
the same
bin we get $\sigma (B-V)~=~0.047 ~mag.$, indicating that an important
fraction of   
the color spread is coming from instrumental errors. The deconvolution
gives a final $\sigma(B-V)~=~0.043 ~mag.$, which is smaller than the
predicted
one. This is still an upper limit because binary stars,
peculiar objects
and field stars contributes in widening the intrinsic distribution.
Similar results are obtained changing the bin width and the position  
along the SGB. \\
Possible explanations to this relatively narrow photometric dispersion
are the following:\\

\begin{description}
\item (1) individual element ratios are effective in
reducing the global $[Me/H]$ spread, or
\item (2) the metal rich component is some
billion years younger than the metal poor one, as discussed by Norris et
al.
(1997).
\end{description}

The first hypothesis implies a trend of decreasing $CNO$ or $s$-elements
with increasing $Fe$, which is
not supported by from Norris and Da Costa
(1995) high resolution analysis. The age effect is an interesting
possibility, with important consequences on the metal enrichment
history of the cluster,
but further work is still needed to check if the observed residual is due   
to some systematical effects in the color transformations or in other
ingredients used in the theoretical models (Cayrel et al., 1997).
A further test could be the measurement of the MS intrinsic width
(see also Noble et al., 1991, Bell and Gustafsson, 1983).
Unfortunately the color-metallicity
relationship in the MS is expected to give a color spread
about
twice smaller than in the SGB. Our tests performed
from the best photometry at about $1 ~mag.$
below the turnoff show that the intrinsic width is completely hidden by
the binary star sequence.\\

\acknowledgments
We would like to thank Roger Cayrel and Alessandro Bressan
for suggestions and helpful discussions.


\begin{references}
\reference Bell R. A., Gustafsson B., 1983, MNRAS 204, 249
\reference Bertelli G., Bressan A., Chiosi C., Fagotto F., Nasi E. 1994,
A\&AS 106, 275
\reference Cayrel R., Lebreton Y., Perrin M-N., Turon C., 1997,
Proceedings of the ESA Symposium `Hipparcos - Venice '97',
13-16 May, Venice, Italy, ESA SP-402 (July 1997), p. 219-224
\reference Da Costa G. S., Villumsen J. E., 1981 in {\it Astrophysical
Parameters for Globular Clusters}, A.G.D. Philips and D.S. Hayes eds., p.
527 
\reference Noble R. G., Dickens R. J., Buttress J., Griffiths W. K., Penny
A. J., 1991, MNRAS 250, 314
\reference Norris J. E., Da Costa G. S., 1995, \apj  447, 680
\reference Norris J. E., Freeman K. C., Mayor M., Seitzer P., 1997, \apj
487, L187
\reference Norris J. E., Freeman K. C., Mighell K. J., 1996, \apj 462,
241
\reference Suntzeff N. B., Kraft R. P., 1996, AJ 111, 1913
\reference Wolley R. v. d. R., 1966, R. Obs. Ann., No. 2.
\end{references}
\end{document}